\pdfoutput=1
\documentclass[journal]{IEEEtran}
\usepackage{amsmath,amsfonts,amssymb}
\usepackage{cite}
\usepackage{graphicx}
\usepackage{subfigure}
\usepackage{url}
\begin{document}
\title{ICWLM\@: A Multi-Task Wireless Large Model\\ via In-Context Learning}
\author{\IEEEauthorblockN{
Yuxuan Wen, Xiaoming Chen, Maojun Zhang, Zhaohui Yang, Chongwen Huang, and Zhaoyang Zhang}
\thanks{Yuxuan Wen, Xiaoming Chen, Maojun Zhang, Zhaohui Yang, Chongwen Huang, and Zhaoyang Zhang are with the College of Information Science and Electronic Engineering, Zhejiang University, Hangzhou, 310027, China (e-mail:\{yuxuanwen, chen\_xiaoming\, zhmj, yang\_zhaohui, chongwenhuang, ning\_ming\}@zju.edu.cn).}
}\maketitle

\begin{abstract}
The rapid evolution of wireless communication technologies, particularly massive multiple-input multiple-output (mMIMO) and millimeter-wave (mmWave), introduces significant network complexity and computational demands. 
Significant research efforts have been made to improve physical layer performance by resorting to deep learning (DL) methods, which, however, are usually task-specific and struggle with data scarcity and generalization.
To address these challenges, we propose a novel In-Context Wireless Large Model (ICWLM), a wireless-native foundation model designed for simultaneous multi-task learning at the physical layer. 
Unlike conventional methods that adapt wireless data to pre-trained large language models (LLMs), ICWLM is trained directly on large-scale, mixed wireless datasets from scratch. 
It jointly solves multiple classical physical layer problems, including multi-user precoding (sum-rate maximization and max-min SINR) and channel prediction. 
A key innovation of ICWLM is its utilization of in-context learning (ICL), enabling the model to adapt to varying system configurations and channel conditions with minimal demonstration pairs, eliminating the need for extensive retraining. 
Extensive simulation results demonstrate that ICWLM achieves competitive performance compared to task-specific methods while exhibiting remarkable generalization capabilities to unseen system configurations. 
This work offers a promising paradigm for developing unified and adaptive AI models for future wireless networks, potentially reducing deployment complexity and enhancing intelligent resource management.
\end{abstract}


\section{Introduction}\label{sec:intro}
With the emergence of advanced wireless communication technologies, such as millimeter-wave (mmWave) and multiple-input multiple-output (MIMO), the complexity of network design has increased significantly~\cite{mMIMO1,mMIMO2,mMIMO3}. 
While enabling unprecedented data throughput and contributing to reduced latency, these technologies come at the cost of large-scale antenna arrays and computationally intensive signal processing strategies.
To overcome these limitations, there has been a growing interest in integrating artificial intelligence (AI) techniques into wireless communication systems~\cite{6G-Net}. 
Unlike traditional model-driven approaches, AI-based methods especially deep learning (DL) are data-driven and capable of learning complex mappings directly from observed data, making them particularly well-suited for environments that are highly dynamic, nonlinear, or difficult to model analytically~\cite{DLinPL}.
In line with this, several white papers in Sixth-Generation (6G) wireless communications envision AI-native wireless networks as a foundational paradigm, where AI is deeply embedded across the communication stack to enable intelligent, self-optimizing, and adaptive systems~\cite{Ali2020,Samsung2022,ATIS2023}.

\begin{figure*}
    \centering
    \includegraphics[width=1.0\linewidth]{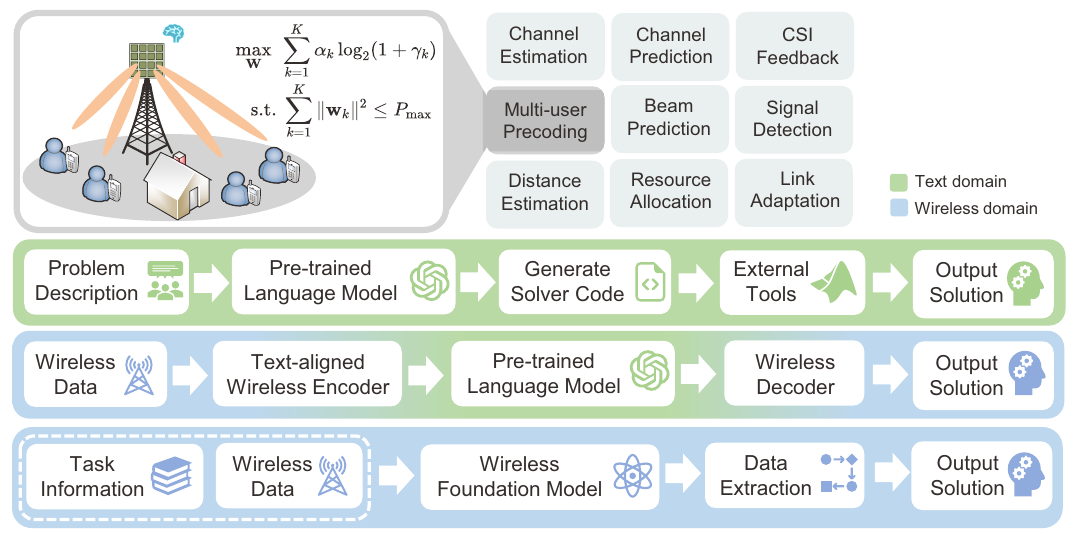}
    \caption{Comparison of different wireless large models.}\label{fig:wllm}
\end{figure*} 

Recent years have witnessed a growing interest in applying DL techniques to physical layer tasks in wireless communications. 
In the area of channel prediction, the work in~\cite{DTL-CP} utilized a fully connected neural network within a deep transfer learning framework to transfer knowledge from uplink to downlink domains, further enhanced by meta-learning for adaptation in low-data scenarios. 
In~\cite{T-CP}, a transformer-based model was introduced to exploit attention mechanisms for parallel multi-frame channel state information (CSI) prediction, effectively addressing channel aging and error propagation in mobile environments. 
Regarding precoding, the authors in~\cite{xia-BF} developed beamforming neural networks (BNNs) based on convolutional neural networks (CNNs) to solve multiple-input single-output (MISO) downlink optimization problems, integrating expert knowledge to ensure near-optimal performance with reduced complexity. 
The study in~\cite{zhang-BF} proposed a customized deep neural network (DNN) for low-complexity precoding in multi-user multiple-input multiple-output orthogonal frequency-division multiplexing (MU-MIMO-OFDM) systems, employing input reduction and model compression techniques to achieve weighted minimum mean square error (WMMSE) level performance with lower computational cost. 
These studies demonstrate the potential of DL models in tackling core physical layer challenges efficiently and effectively.
Though a lot of research efforts have been made, there still exist some critical issues that restrain their practical applications.
The first challenge lies in the difficulty of obtaining sufficient high-quality labeled data, especially in dynamic and complex wireless environments. 
The performance of DL models often depends heavily on network size and supervised training, which becomes a bottleneck when annotated data is scarce or expensive to acquire. 
The second challenge concerns the limited generalization capability of existing models. 
Since most DL architectures are trained for specific scenarios or channel conditions, they require frequent retraining or fine-tuning to adapt to evolving environments, 
leading to increased computational overhead and latency. 
Finally, most current DL-based methods are task-specific, designed separately for individual problems such as channel estimation, prediction, or precoding. 
The fragmented design additionally introduces significant deployment cost as the base station needs to save all of the corresponding models. 
Given this, there is an urgent need for a unified framework capable of addressing multiple physical layer tasks in a flexible manner, while also being robust to dynamic channel conditions and system configurations. 

More recently, large language models (LLMs) such as ChatGPT~\cite{ChatGPT}, LLaMa~\cite{LlaMa}, and DeepSeek~\cite{DeepSeek-R1} have gained significant attention and achieved remarkable success in various fields. 
These models, characterized by their massive parameter sets, have demonstrated exceptional performance in multiple natural language processing (NLP) tasks, including language modeling, translation, and question-answering, among others.
Unlike smaller models, large models (LMs) can capture extensive, universal knowledge and demonstrate outstanding performance in various tasks. 
Their ability to perform multiple tasks with minimal fine-tuning has made them powerful tools for various domains, including computer vision~\cite{cv}, robotics~\cite{robotics}, and healthcare~\cite{healthcare}.
Recent advancements in LLMs have also shown great promise in transforming various aspects of wireless communications, particularly in the physical layer tasks. 
WirelessLLM~\cite{WirelessLLM}, the pioneering work, has specifically adapted and enhanced LLMs for wireless communication systems by employing techniques such as knowledge alignment, fusion, and evolution.
The performance of WirelessLLM in case studies, including power allocation, spectrum sensing, and protocol understanding, demonstrates the practical applicability of this model in addressing key problems within wireless networks. 
Subsequently, the works in~\cite{LLM4CP} and~\cite{CsiLLM} have proposed pre-trained LLM-empowered methods for channel prediction.
To facilitate effective cross-modality applications, these methods tailor specific modules to bridge the gap between raw wireless communication data and the feature space of pre-trained LLM\@.
While freezing most of its parameters, the LLM is fine-tuned on corresponding CSI datasets, enabling it to adapt to the unique characteristics of wireless communication tasks.
Besides the aforementioned works, LLMs have been applied to various tasks, such as beam prediction~\cite{LLM4BF}, channel feedback~\cite{LLM4FB1, LLM4FB2}, and resource scheduling~\cite{LLM4RA1, LLM4RA2}. 
In the latest research, several works have explored the potential of using a single LLM to solve multiple tasks, leveraging the inherent correlations between them. 
In~\cite{Multi-taskLLM}, the authors proposed an LLM-enabled multi-task physical layer network that unified multiple tasks within a single LLM\@. 
This approach employed task-specific encoders and decoders to handle the input and output data, and then fine-tuned the LLM backbone across a diverse range of tasks, including multi-user precoding, channel prediction, and signal detection.
The proposed model outperformed traditional task-specific approaches, highlighting the capability of LLMs to efficiently handle multiple physical layer tasks in parallel. 
Similarly, LLM4WM~\cite{LLM4WM} presented a customized framework for channel-related tasks, achieving superior results over existing methods in both full-sample and few-shot evaluation settings.
These advancements harnessed the potential of LLMs to extract shared channel representations, providing a novel solution for deploying large models in real-world wireless communication environments.
Despite the promising progress outlined above, existing efforts that integrated LLMs into wireless communications fundamentally relied on aligning or adapting wireless-domain data to the textual input space of pre-trained LLMs. 
This cross-domain adaptation process, which typically involves feature engineering, prompt formatting, or intermediate representation mapping, is inherently indirect and comes with several limitations. 
These include reduced representation fidelity, increased data preprocessing complexity, and limited scalability when applied to new or diverse wireless tasks.

\begin{figure}
    \centering
    \includegraphics[width=1\linewidth]{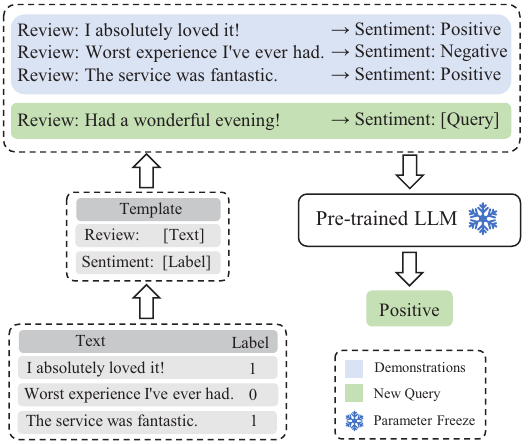}
    \caption{Illustration of In-Context Learning (ICL) mechanism.}\label{fig:icl}
\end{figure}

To address these challenges, we propose a paradigm shift as illustrated in Fig.~\ref{fig:wllm}.
Specifically, rather than adapting wireless data to fit to the structure of pre-trained LLMs, we envision a foundation model that is inherently designed for the wireless communication domain. 
This model is trained directly on large-scale wireless data, without requiring transformation into the language modality. 
To enable efficient multi-task learning within a wireless-native foundation model, a swift and adaptable learning mechanism is essential. 
In-context learning (ICL), an emerging capability from large language models, offers a promising solution.
As shown in Fig.~\ref{fig:icl}, ICL enables a model to adapt to new tasks with minimal data by learning directly from a few examples provided in the input, without the need for parameter updates or additional training~\cite{ICL}. 
ICL adaptability is particularly promising for wireless communications, where system configurations, network conditions, and channel environments change frequently. 
Building upon these capabilities, we aim to extend the application of ICL to a broader range of physical layer tasks in wireless communications. 
By leveraging ICL, a unified framework can be developed to address challenges such as channel prediction and multi-user precoding simultaneously.
The main contributions are listed as follows:
    \begin{itemize}
        \item We propose a novel In-Context Wireless Large Model (ICWLM) for multiple classical physical layer tasks in wireless communications. Leveraging the ICL mechanism, the proposed model is designed to jointly solve multi-user precoding and channel prediction tasks, adapting to different system configurations and channel conditions with minimal demonstration pairs.
        \item To realize a wireless-native foundation model, we design a causal transformer backbone capable of processing complex-valued wireless data, which is formatted into unified input-output sample pair sequences for efficient processing. This model is trained from scratch on a large-scale, mixed dataset from diverse physical layer scenarios using a self-supervised approach. 
        \item Extensive simulation results demonstrate that ICWLM achieves competitive performance compared to task-specific methods. Crucially, it exhibits remarkable generalization capabilities to unseen system configurations (e.g., varying SNR values) with only a few in-context demonstration pairs, significantly reducing the need for extensive retraining.
    \end{itemize}

The remainder of the paper is organized as follows. 
Section~\ref{sec:system_model} introduces the system model and formulates the multi-user precoding and channel prediction problems.
Section~\ref{sec:approach} presents the proposed ICWLM model, including the model architecture, data formulation, and multi-task training schedule.
Section~\ref{sec:simulations} provides simulation results to evaluate the performance of the proposed model.
Finally, Section~\ref{sec:conclusion} concludes the paper and discusses future research directions. 

\textbf{Notations:} 
The notations are given as follows. 
Matrices and vectors are denoted by bold capital and lowercase symbols. 
${(\mathbf{A})}^T$ and ${(\mathbf{A})}^H$ stand for transpose and conjugate transpose of $\mathbf{A}$, respectively.
The notations $||\bullet||_F$ represents the Frobenius norm.
$\otimes$ and $\odot$ denotes the Kronecker product and Hadamard product, respectively.
$\mathcal{CN}(\mu,\sigma^2)$ denotes the complex Gaussian distribution with mean $\mu$ and variance $\sigma^2$.

\begin{figure}
    \centering
    \includegraphics[width=0.9\linewidth]{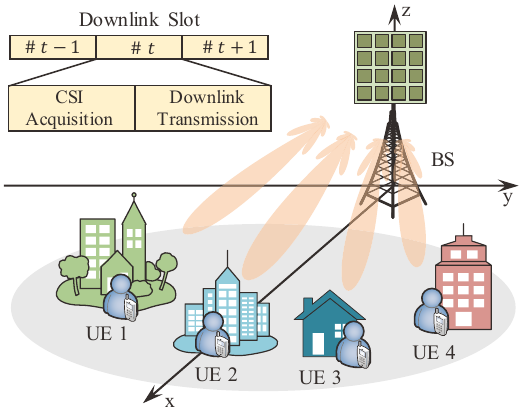}
    \caption{Schematic diagram of multi-user downlink transmission system.}\label{fig:system}
\end{figure}

\section{System Model}\label{sec:system_model}
For simplicity but without loss of generality, we consider a multi-user multiple-input signle-output (MU-MISO) system, 
where a base station (BS) equipped with $N_t$ antennas serves $K$ single-antenna users simultaneously. 
The BS is equipped with a uniform planar array (UPA) consisting of $N_t=N_h \times N_v$ antennas, 
where $N_h$ and $N_v$ denote the number of antennas in the horizontal and vertical directions, respectively. 
The system works in slotted time as shown in Fig.~\ref{fig:system}. A time slot could be divided into two phases, 
namely the CSI acquisition phase and the downlink data transmission phase.
We assume that downlink transmission occurs over quasi-static block fading channel,
i.e., the channel remains time-invariant in each slot and changes from slot to slot.

\subsection{Channel Model}
\begin{figure}
    \centering
    \includegraphics[width=1\linewidth]{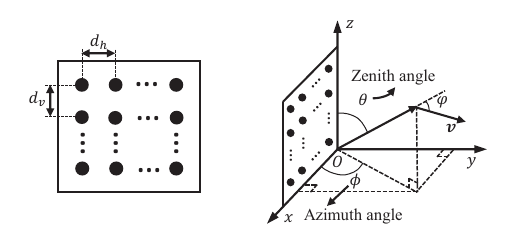}
    \caption{UPA antenna configuration in 3D-Cartesian coordinate system.}\label{fig:antenna}
\end{figure}

In this work, we adopt the cluster-based multi-path channel model in~\cite{3gpp901}. 
The downlink channel $\mathbf{h}\in \mathbb{C}^{N_t \times 1}$ between the BS and the user is characterized at certain time $t$ and frequency $f$ as follows:
\begin{equation}\label{CSI}
\mathbf{h}(t,f)=\sum_{c=1}^{N_c} \sum_{p=1}^{N_p}g_{c,p}e^{j[2\pi (w_{c,p} t-f\tau_{c,p})+\Phi_{c,p}]}\mathbf{a}(\theta_{c,p},\phi_{c,p}),
\end{equation}
where $N_c$ and $N_p$ represent the number of clusters and the number of paths within each cluster. The corresponding parameters of each path $p$ in each cluster $c$ are the complex path gain $g_{c,p}$, Doppler frequency shift $w_{c,p}$, delay $\tau_{c,p}$, random phase $\Phi_{c,p}$, and steering vector $\mathbf{a}(\theta_{c,p},\phi_{c,p})\in \mathbb{C}^{N_t \times 1}$.
The downlink Doppler frequency shift is defined as $w_{c,p}=v\cos(\varphi_{c,p})f/c$, where $v$ is the user velocity, $\varphi_{c,p}$ is the angle between the direction of velocity vector and the path, $f$ is the carrier frequency, and $c$ is the speed of light.
Denote the zenith angle and azimuth angle by $\theta_{c,p}$ and $\phi_{c,p}$, respectively.
Fig.~\ref{fig:antenna} demonstrates the UPA antenna configuration in 3D-Cartesian coordinate system.
The steering vector of the corresponding path $\mathbf{a}(\theta_{c,p}, \phi_{c,p})$ is modeled as the Kronecker product of the vertical steering vector $\mathbf{a}_{v}(\theta_{c,p})\in \mathbb{C}^{N_v \times 1}$ and horizontal steering vector $\mathbf{a}_{h}(\theta_{c,p}, \phi_{c,p})\in \mathbb{C}^{N_h \times 1}$
\begin{equation}
    \mathbf{a}(\theta_{c,p},\phi_{c,p})=\mathbf{a}_{h}(\theta_{c,p},\mathbf{\phi}_{c,p})\otimes\mathbf{a}_{v}(\theta_{c,p}),
\end{equation}
where 
\begin{align}
\mathbf{a}_{h}\left(\theta_{c,p},\phi_{c,p}\right) & =
\begin{bmatrix}
1 \\
e^{j2\pi\frac{d_{h}f}{c}\cos\theta_{c,p}\cos\phi_{c,p}} \\
\vdots \\
e^{j2\pi\frac{d_{h}f}{c}(N_{h}-1)\cos\theta_{c,p}\cos\phi_{c,p}}
\end{bmatrix},\quad &\\
\mathbf{a}_{v}\left(\theta_{c,p}\right) & =
\begin{bmatrix}
1 \\
e^{j2\pi\frac{d_{v}f}{c}\cos\theta_{c,p}} \\
\vdots \\
e^{j2\pi\frac{d_{v}f}{c}(N_{v}-1)\cos\theta_{c,p}}
\end{bmatrix},\quad &
\end{align}
and $d_h$ and $d_v$ denote the antenna spacing in the horizontal and vertical directions.

\subsection{Multi-user Precoding}
For a downlink transmission scenario, the channel between user $k$ and the BS is denoted as ${\bf h}_k \in \mathbb{C}^{N_t \times 1}, k=1,2,\ldots,K$. 
The received signal at user $k$ is given by
\begin{equation}\label{eq:signal}
    y_{k} = \mathbf{h}_{k}^{H} \sum_{i=1}^{K} \mathbf{w}_{i} x_{i} + n_{k},
\end{equation}
where $\mathbf{w}_{i}$ represents the precoding vector for user $i$, $x_{i} \sim \mathcal{CN}(0,1)$ is the transmitted symbol from the BS to user $i$, 
and $n_{i} \sim \mathcal{CN}(0,\sigma^2)$ denotes the additive white Gaussian noise (AWGN) with zero mean and variance $\sigma^2$.
Based on the signal model~\eqref{eq:signal}, the received signal to interference plus noise ratio (SINR) of user $k$ equals to
\begin{equation}\label{eq:sinr}
  \gamma_k=\frac{|\mathbf{h}^H_k\mathbf{w}_k|^2}{\sum^K_{i=1,i\neq k}|\mathbf{h}^H_k\mathbf{w}_{i}|^2+\sigma^2}.
\end{equation}

\subsubsection{Sum-Rate Maximization Problem}
The first task aims to optimize the precoding vectors
such that the system sum-rate is maximized subject to a total power constraint due to the BS power budget. 
For simplicity, we design the precoding vectors based on the channel of the central carrier-frequency ${\bf h}_k$ 
and the problem is mathematically formulated as
\begin{subequations}\label{eq:p1}
    \begin{align}
    \textbf{P1: } \max_{\mathbf{W}}\quad & \sum^{K}_{k=1}\alpha_{k}\log_{2}(1+\gamma_{k}), \label{eq:p1a}\\
    \text{s.t.}\quad & \sum^K_{k=1}\|\mathbf{w}_k\|^2\leq P_{\text{max}}, \label{eq:p1b}
    \end{align}
\end{subequations}
where $\alpha_k$ represents the priority weight for user $k$, $\mathbf{W}=[\mathbf{w}_{1},\mathbf{w}_{2},\ldots,\mathbf{w}_{K}] \in \mathbb{C}^{N_t \times K}$
is a set of precoding vectors and $P_{\text{max}}$ is the power budget.
Direct prediction of the high-dimensional matrix $\mathbf{W}$ via neural networks is often suboptimal due to the large solution space.
However, based on the optimal beamforming structure pointed out in~\cite{optimal}, the optimal precoder $\mathbf{w}_k^*$ typically follows a structure as
\begin{equation}\label{eq:optimal_structure_p1}
    \mathbf{w}_k^* = \sqrt{p_k} \frac{{\left( \mathbf{I}_{N_t} + \sum_{k=1}^{K} \frac{\lambda_k}{\sigma^2} \mathbf{h}_j \mathbf{h}_j^H \right)}^{-1} \mathbf{h}_k}{\left\| {\left( \mathbf{I}_{N_t} + \sum_{k=1}^{K} \frac{\lambda_k}{\sigma^2} \mathbf{h}_j \mathbf{h}_j^H \right)}^{-1} \mathbf{h}_k \right\|},
\end{equation}
where $p_k$ denotes the allocated downlink power for user $k$, $\lambda_k$ is a positive parameter and $\sum_{k=1}^{K}p_k =\sum_{k=1}^{K}\lambda_k=P_{\text{max}}$.
The solution structure reduces the dimension of the solution space from $N_t \times K$ complex variables to $2K$ real variables, significantly simplifying the learning task.

\subsubsection{SINR Balancing Problem}
The second task addresses the fairness issue by maximizing the minimum SINR among all users, which is formulated as
\begin{subequations}\label{eq:p2}
    \begin{align}
    \textbf{P2: }\max_{\mathbf{W}}\min_{1\leq k\leq K}\quad & \frac{\gamma_k}{\rho_k}, \label{eq:p2a}\\
    \text{s.t.}\quad & \sum^K_{k=1}||\mathbf{w}_k||^2\leq P_{\text{max}}, \label{eq:p2b}
    \end{align}
\end{subequations}
where $\rho_k$ are constant weights denoting the importance of the user $k$.
Such an optimization problem is referred to as interference or SINR balancing, and has been investigated in many works~\cite{Schubert-Boche, 1561584, 4203115}.
According to the uplink-downlink duality theory~\cite{Schubert-Boche}, 
The normalized beamforming vectors in problem P2 also follow a similar structure as
\begin{equation}\label{eq:optimal_structure_p2_direction}
    \bar{\mathbf{w}}_k^* = \frac{{\left(\mathbf{I}_{N_t} + \sum_{j=1}^{K} \frac{q_j}{\sigma^2} \mathbf{h}_j \mathbf{h}_j^H \right)}^{-1} \mathbf{h}_k}{\left\| {\left( \mathbf{I}_{N_t} + \sum_{j=1}^{K} \frac{q_j}{\sigma^2} \mathbf{h}_j \mathbf{h}_j^H \right)}^{-1} \mathbf{h}_k \right\|},
\end{equation}
where $q_j$ denotes the virtual uplink power for user $j$.
Let $\mathbf{q} = {[q_1, \dots, q_K]}^T$ denote the virtual uplink power allocation vector. 
Once the directions are fixed via $\mathbf{q}$, the optimal downlink power vector $\mathbf{p}$ can be uniquely determined by an algorithm in~\cite[Table 1]{Schubert-Boche}.
Thus, the optimal precoding vectors can be obtained as
\begin{equation}\label{eq:optimal_structure_p2_final}
   \mathbf{w}_k^* = \sqrt{p_k} \bar{\mathbf{w}}_k^*.
\end{equation}
Instead of predicting $\mathbf{W}$ directly, we can predict the uplink power allocation vector $\mathbf{q}$, which reduces the output dimension from $N_t \times K$ complex variables to $K$ real variables.

\subsection{Channel Prediction}
Precoding in downlink transmission using timely CSI can achieve a satisfactory performance in stationary scenarios.
However, the obtained CSI is often outdated due to user mobility and feedback delay, 
which may lead to performance degradation for precoding design.
To mitigate the channel aging issue, 
accurate channel prediction is essential to predict the current downlink CSI based on historical CSI\@.
With the predicted CSI, the BS can optimize the precoding vectors to improve the system performance.

In practical communication scenarios, the mobility of cellular users introduces Doppler frequency shifts, leading to variations in CSI\@. 
In TDD systems, due to channel reciprocity, the downlink CSI can be obtained at the BS side by channel estimation on uplink pilots.
While in FDD systems where the frequency of the uplink and downlink channels differs, 
downlink CSI can only be estimated at the user side and then fed back to the BS\@. 
The CSI estimation and feedback process incur additional computational and transmission time overhead, 
causing channel aging in high dynamic scenarios as previously described. 
Therefore, channel prediction at the BS side becomes essential to mitigate the impact of latency and errors in downlink transmission.

We consider a time-varying massive MIMO channel assuming that the variation is caused by the user mobility while the BS is static.
The channel between the BS and user $k$ at time slot $t$ is denoted as $\mathbf{h}_k^t \in \mathbb{C}^{N_t \times 1}$. 
For simplicity, we organize all users' CSI into a matrix form as
\begin{equation}
    \mathbf{H}^t=[\mathbf{h}_1^t,\mathbf{h}_2^t,\ldots,\mathbf{h}_K^t]\in \mathbb{C}^{N_t \times K},
\end{equation}
where $\mathbf{H}^t$ is the downlink CSI matrix of all users at time slot $t$. 
In this work, we focus on the one-step channel prediction problem to address the issue of outdated CSI\@. 
We aim to accurately predict CSI of current time slot $t$ based on historical CSI of $T$ time slots.
The normalized mean square error (NMSE) between predicted and actual downlink CSI is used to evaluate the prediction accuracy. 
Then the entire problem can be described as follows:
\begin{subequations}\label{eq:p3}
    \begin{align}
    \textbf{P3: } \min_{\Omega} \quad &  \frac{ \| \hat{\mathbf{H}}^{t} - \mathbf{H}^{t} \|_F^2}{\| \mathbf{H}^{t} \|_F^2}, \label{eq:p3a}\\
    \text{s.t.} \quad &  \hat{\mathbf{H}}^{t} = f_{\Omega}(\mathbf{H}^{t-1},\mathbf{H}^{t-2},\ldots, \mathbf{H}^{t-T}),\label{eq:p3b}
    \end{align}
\end{subequations}
where $\hat{\mathbf{H}}^{t}$ represents the predicted CSI, and $\mathbf{H}^{t}$ denotes the ground-truth downlink CSI at time slot $t$. $f_{\Omega}$ is the mapping function with trainable parameters $\Omega$.
It is worth noting that in the simulation phase, $\{\mathbf{H}^{t}, \mathbf{H}^{t-1},\ldots, \mathbf{H}^{t-T}\}$ are obtained from the channel generator as the ground truth label.
While in practical systems, the BS can acquire $\{\mathbf{H}^{t-1},\ldots, \mathbf{H}^{t-T}\}$ through channel feedback from users, and then predict $\hat{\mathbf{H}}^{t}$ for downlink transmission.

\section{Proposed Multi-Task Wireless Large Model}\label{sec:approach}
In this section, we introduce the proposed In-Context Wireless Large Model (ICWLM), which is designed to address several physical layer tasks in wireless communication systems. 
To achieve this, we leverage the powerful ICL paradigm, enabling our model to handle multiple tasks within a single unified framework. 
This approach eliminates the need for task-specific architecture modifications, enabling the model to generalize effectively to unseen configurations. 

\subsection{In-Context Learning}\label{sec:icl}
When training a single unified model for multiple different tasks, the main challenge lies in identifying and adapting to the specific task at hand without explicit supervision or retraining. 
Fortunately, ICL offers a promising paradigm by enabling the model to infer task intent directly from a sequence of task-specific input-output examples embedded in the input, rather than relying on task-specific architectural branches or metadata.
In the case of wireless communication systems, this context is constructed from structured communication data—such as channel matrices, received signals, or pilot sequences—allowing the model to distinguish tasks and generalize to new configurations during inference.

\begin{figure}
    \centering
    \includegraphics[width=1\linewidth]{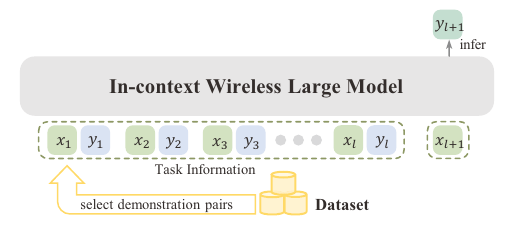}
    \caption{The proposed ICL framework for multi-task wireless large model.}\label{fig:framework}
\end{figure}

Based on the definition in~\cite{Bhasin2024HowDM}, we now concretely describe a general methodology for training a wireless large model that can in-context learn a specific task.
The core idea of this methodology is to leverage the ICL capability of transformer-based large models to adapt to diverse wireless tasks without requiring task-specific retraining. 
Therefore, the ICL problem could be formulated as passing in an $l$-shot sequence $\mathcal{S}^{l}=\{x_1,f(x_1), x_2,f(x_2), \dots, x_l,f(x_l), x_{l+1}\}$ to the model $M_{\theta}$
and generating an output $M_{\theta}(\mathcal{S}^{l})$ to predict the ground-truth $f(x_{l+1})$, where the examples have not been seen during training.
Here, the function $f$ is task-dependent and varies across different physical-layer applications as described in Section~\ref{sec:system_model}.
For instance, if $x$ represents the CSI matrix, then in the precoding task $f(x)$ denotes the precoding matrix computed for a given channel.
Similarly, in the channel prediction task, $f(x)$ represents the channel state at the current time slot conditioned on past observations.
The training objective is to minimize the expected loss over all input-output pairs, which can be expressed as
\begin{align}\label{eq:icl_objective}
    \min_{\theta} \mathbb{E}_{\mathcal{S}} \left[ \frac{1}{l+1} \sum_{i=0}^{l} \ell\left(M_{\theta}\left(\mathcal{S}^{i}\right), f\left(x_{i+1}\right)\right) \right],
\end{align}
where $\ell(\cdot, \cdot)$ is an appropriately chosen loss function, such as the mean squared error (MSE) or nomarlized mean squared error (NMSE) for regression tasks.
During the inference phase, the model utilizes a small number of demonstration pairs as prompts to infer the task intent and generate the desired output. 
The proposed framework is illustrated in Fig.~\ref{fig:framework},
which enables the model to generalize to new configurations with minimal examples, making it particularly suitable for complex wireless environments.

\subsection{Data Formulation}\label{sec:data}
To fully harness the ICL capability of our model, the first step is to appropriately format the task-specific wireless datasets into a sequence of input-output token pairs.
This data reorganization is essential for leveraging the sequential nature of transformer models while preserving the inherent structure of wireless communication data.

For both precoding and channel prediction tasks, the raw data consists of complex-valued channel matrices $\mathbf{X} \in \mathbb{C}^{N_t \times K}$, 
where $N_t$ represents the number of transmit antennas at the BS and $K$ denotes the number of users.
Since neural networks generally deal with real numbers, we first decompose each complex matrix into its real and imaginary parts, resulting in 
\begin{equation}
    \mathbf{X}_\text{real} = \text{Re}(\mathbf{X}), \quad \mathbf{X}_\text{imag} = \text{Im}(\mathbf{X}),
\end{equation}
The real and imaginary components $\mathbf{X}_\text{real}\in \mathbb{C}^{N_t \times K}$ and $\mathbf{X}_\text{imag}\in \mathbb{C}^{N_t \times K}$ are then vectorized and concatenated separately,
yielding a real-valued vector $\mathbf{x} \in \mathbb{R}^{2N_{t}K}$. 

After processing the raw wireless data and organizing it into the appropriate format, the ICL sequences for specific tasks can be constructed as follows.
For the precoding task, the $l$-shot sequence could be constructed straightly as
\begin{equation}
    \mathcal{S}^{l}_{\text{precoding}}=\{\mathbf{x}_1,\mathbf{y}_1,\mathbf{x}_2,\mathbf{y}_2,\ldots,\mathbf{x}_l,\mathbf{y}_l,\mathbf{x}_{l+1}\},
\end{equation}
where $\mathbf{x}_i$ represents the processed CSI and $\mathbf{y}_i$ contains the corresponding low-dimensional parameter vectors that determine the optimal precoder.
Specifically, for the sum-rate maximization task (P1), $\mathbf{y}_i$ is a concatenation of the power allocation vector $\mathbf{p}$ and the dual variable vector $\mathbf{\lambda}$, 
while for the SINR balancing task (P2), it contains the virtual uplink power vector $\mathbf{q}$. 
To ensure that all tokens within the ICL sequence have a uniform dimension, the low-dimensional parameter vectors forming $\mathbf{y}_i$ are padded with zeros to match the dimension of the input vector $\mathbf{x} \in \mathbb{R}^{2N_{t}K}$.
This unified token size is essential for the transformer architecture to process the input-output pairs in a consistent manner.
For the channel prediction task, to capture the temporal dependencies in the data, we employ an overlapping arrangement of the CSI tokens.
Specifically, the $l$-shot sequence for this task is constructed as
\begin{equation}
    \mathcal{S}^{l}_{\text{prediction}}=\{\mathbf{x}_1,\mathbf{x}_2,\mathbf{x}_2, \mathbf{x}_3,\ldots,\mathbf{x}_l,\mathbf{x}_{l+1},\mathbf{x}_{l+1}\},
\end{equation}
where each pair of consecutive CSI tokens $(\mathbf{x}_{i},\mathbf{x}_{i+1})$ captures the relationship between adjacent channel states. 
This arrangement enables the model to learn the temporal correlation between adjacent channel states effectively,
while maintaining the consistent input-output pair structure required by the in-context learning framework.

\subsection{Model Architecture}
\begin{figure}
    \centering
    \includegraphics[width=1\linewidth]{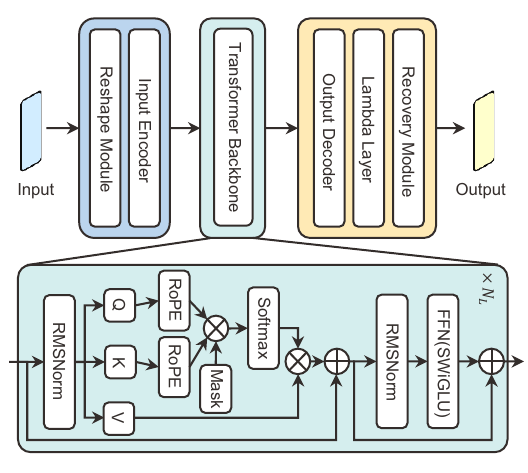}
    \caption{The proposed ICWLM model architecture.}\label{fig:model}
\end{figure}

We construct the proposed ICWLM model with three parts: an input module, a causal transformer backbone, and an output module.
The model architecture is shown in Fig.~\ref{fig:model}.

\subsubsection{Input Module}
The input module is specifically engineered to project high-dimensional wireless entities, such as channel matrices and precoding parameters, into a tokenized latent space compatible with the decoder-only transformer architecture.
To achieve this, we reshape the input matrices as mentioned in Section~\ref{sec:data}
and use a linear encoder to transform the raw input data into a latent representation that aligns with the transformer model's input format, 
ensuring compatibility and efficient processing.

For example, given an input vector $\mathbf{x} \in \mathbb{R}^{1\times2N_{t}K}$, 
which represents the real and imaginary parts of the wireless data concatenated into a single vector, 
the encoder applies a linear transformation to project the input into a higher-dimensional latent space. 
This transformation is expressed as
\begin{equation}
    \mathbf{x}_e = \mathbf{x}\mathbf{W}_e + \mathbf{b}_e,
\end{equation}
where $\mathbf{W}_e \in \mathbb{R}^{2N_{t}K\times d}$ is the weight matrix, $\mathbf{b}_e \in \mathbb{R}^{1\times d}$ is the bias vector, and $d$ denotes the embedding dimension of the model. 
Both $\mathbf{W}_e$ and $\mathbf{b}_e$ are learnable parameters optimized during training.
This design ensures that the input module effectively captures the essential features of the wireless communication data while maintaining a lightweight structure. 
By avoiding overly complex architectures, the linear input module achieves a balance between computational efficiency and representation power, enabling seamless integration with the subsequent transformer layers. 

\subsubsection{Transformer Backbone}
The encoded sequences of channel matrices and precoding parameters are fed into the transformer backbone, where we adopt the LLaMA architecture~\cite{LlaMa}. 
As a state-of-the-art decoder-only language model, LLaMA incorporates several critical architectural improvements over the standard Transformer, making it exceptionally robust for learning complex representations from high-dimensional wireless data.
The backbone consists of $N_L$ stacked decoder layers, each integrating three key components: Root Mean Square Normalization (RMSNorm)~\cite{rmsnorm}, Multi-Head Self-Attention (MHSA)~\cite{attention} with Rotary Position Embeddings (RoPE)~\cite{rope}, and a Feed-Forward Network (FFN) with SwiGLU activation.

To enhance training stability, LLaMA backbone employ RMSNorm for pre-normalization at the input of each sub-layer, rather than the LayerNorm used in the original Transformer.
Given that wireless channel data often exhibits high dynamic ranges due to path loss and fading, RMSNorm effectively stabilizes the gradient by normalizing the input based on the root mean square, facilitating the convergence of deep networks.

For the MHSA, we define the input sequence as $\mathbf{X}_e \in \mathbb{R}^{L \times d}$, where $L$ is the sequence length and $d$ is the hidden dimension. The input is first linearly projected to queries, keys, and values through:
\begin{align}
    \mathbf{Q} &= \mathbf{X}_e\mathbf{W}_Q, \notag\\
    \mathbf{K} &= \mathbf{X}_e\mathbf{W}_K, \notag\\
    \mathbf{V} &= \mathbf{X}_e\mathbf{W}_V,
\end{align}
where $\mathbf{W}_Q, \mathbf{W}_K, \mathbf{W}_V \in \mathbb{R}^{d \times d}$ are learnable projection matrices.
To incorporate positional information, RoPE embeddings are then applied to queries and keys before computing attention scores.
For an embedding vector $\mathbf{x}_e = (x_1, x_2, \ldots, x_{d})$, The RoPE embeddings are computed as
\begin{equation}
\label{eq:your_equation_name_split}
\begin{split}
\mathbf{R}_{\Theta,m}^{d}\mathbf{x}_e^T = & 
    \begin{pmatrix}
    x_1 \\ x_2 \\ x_3 \\ x_4 \\ \vdots \\ x_{d-1} \\ x_d
    \end{pmatrix}
    \odot
    \begin{pmatrix}
    \cos m\theta_1 \\ \cos m\theta_1 \\ \cos m\theta_2 \\ \cos m\theta_2 \\ \vdots \\ \cos m\theta_{d/2} \\ \cos m\theta_{d/2}
    \end{pmatrix} \\
    & +
    \begin{pmatrix}
    -x_2 \\ x_1 \\ -x_4 \\ x_3 \\ \vdots \\ -x_d \\ x_{d-1}
    \end{pmatrix}
    \odot
    \begin{pmatrix}
    \sin m\theta_1 \\ \sin m\theta_1 \\ \sin m\theta_2 \\ \sin m\theta_2 \\ \vdots \\ \sin m\theta_{d/2} \\ \sin m\theta_{d/2}
    \end{pmatrix}
\end{split}
\end{equation}
where $\odot$ denotes the Hadamard product, $m$ is the position of the current token in the input sequence, and $\Theta = \left\{ \theta_i = 10000^{-2(i-1)/d}, \, i \in [1, 2, \dots, d/2] \right\}$ is the set of angles.
This operation ensures that the model can effectively capture the positional relationships between tokens in the sequence, which is critical for tasks involving temporal or spatial dependencies, such as channel prediction in wireless communications.
Let $\text{RoPE}(\mathbf{Q})$ and $\text{RoPE}(\mathbf{K})$ denote the matrices where each row of $\mathbf{Q}$ and $\mathbf{K}$ has been transformed by the RoPE operation according to its position in the sequence.
The attention scores are computed as:
\begin{align}
    &\text{Attention}(\mathbf{Q},\mathbf{K},\mathbf{V}) = \notag\\
    &\quad\text{softmax}{\left(\frac{\text{RoPE}(\mathbf{Q})\text{RoPE}{\left(\mathbf{K}\right)}^T}{\sqrt{d_k}}\right)}\mathbf{V},
\end{align}
where $d_k = d / N_H$ is the dimension of each attention head, with $N_H$ being the number of heads.
The softmax function is applied to each row of the attention score matrix. For a generic input vector $\mathbf{z} = [z_1, \dots, z_L]$, the softmax function is defined as:
\begin{equation}
    \text{softmax}{\left(\mathbf{z}\right)}_i = \frac{e^{z_i}}{\sum_{j=1}^L e^{z_j}}, \quad \text{for } i = 1, \dots, L.
\end{equation}

Following the attention layer, the FFN module processes the hidden states. 
We adopt the SwiGLU activation function~\cite{swiglu} used in LLaMA\@. 
Unlike the standard FFN, SwiGLU utilizes a gating mechanism with three linear projections.
For an input vector $\mathbf{x}_o$, the output is computed as:
\begin{equation}
\text{FFN}(\mathbf{x}_o) = \left(\text{SiLU}\left(\mathbf{x}_o\mathbf{W}_G\right) \odot \left(\mathbf{x}_o\mathbf{W}_U\right)\right)\mathbf{W}_D,
\end{equation}
where $\odot$ denotes Hadamard product, and $\text{SiLU}(x) = x \cdot \text{sigmoid}(x)$. 
The terms $\mathbf{W}_G$, $\mathbf{W}_U$, and $\mathbf{W}_D$ represent the learnable weight matrices for the gate, up-projection, and down-projection, respectively.
This architecture provides a linear path for gradients to propagate during backpropagation, thereby improving the gradient flow compared to pure non-linear functions and facilitating deep network training.
Furthermore, the dimension of these intermediate layers determines the representational capacity of the model, i.e., its ability to approximate complex functions. A sufficiently large dimension enables the model to capture the intricate high-dimensional features inherent in wireless channel data.

\subsubsection{Output Module}
Once the transformer backbone generates the latent representations $\mathbf{h}_{o} \in \mathbb{R}^{1\times d}$, we recover them back to the wireless format through:
\begin{equation}
    \mathbf{y} = \mathbf{h}_o\mathbf{W}_o + \mathbf{b}_o,
\end{equation}
where $\mathbf{W}_o \in \mathbb{R}^{2N_{t}K \times d}$ and $\mathbf{b}_o \in \mathbb{R}^{1\times2N_{t}K}$ are learnable parameters.
For channel prediction tasks, the output vector $\mathbf{y}$ is directly reshaped back to the complex channel matrix format.
For precoding tasks, we extract the relevant low-dimensional parameter vectors from $\mathbf{y}$. To ensure the physical constraint of non-negativity for power allocation and dual variables, a Sigmoid activation function is applied element-wise:
\begin{equation}
    \sigma(z) = \frac{1}{1 + e^{-z}},
\end{equation}
They are then scaled appropriately to satisfy the total power constraints, i.e.,
\[\hat{\mathbf{p}}^* = \frac{P_{\text{max}}}{\|\hat{\mathbf{p}}\|_1} \hat{\mathbf{p}},\quad \hat{\boldsymbol{\lambda}}^* = \frac{P_{\text{max}}}{\|\hat{\boldsymbol{\lambda}}\|_1}\hat{\boldsymbol{\lambda}},\quad \hat{\mathbf{q}}^* = \frac{P_{\text{max}}}{\|\hat{\mathbf{q}}\|_1} \hat{\mathbf{q}}\]
which are then used to reconstruct the optimal precoding matrix based on the structures defined in Eqs.~\eqref{eq:optimal_structure_p1} and~\eqref{eq:optimal_structure_p2_final}.

\subsection{Multi-task Training Schedule}

\begin{figure}
    \centering
    \subfigure[Multi-user Precoding]{\includegraphics[width=1\linewidth]{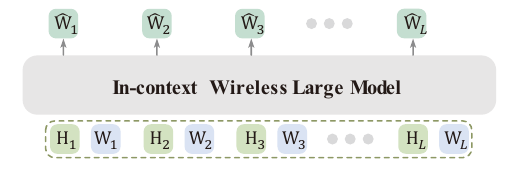}}
    \subfigure[Channel Prediction]{\includegraphics[width=1\linewidth]{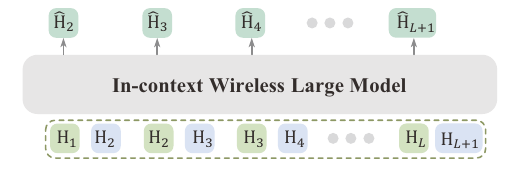}}
    \caption{The proposed training schedule for multi-task learning.}\label{fig:training}
\end{figure}

In this work, we adopt a multi-task training schedule that explicitly leverages ICL to handle multiple physical layer tasks without additional fine-tuning for each task. 
As illustrated in Fig.~\ref{fig:training}, we integrate the precoding (P1, P2) and channel prediction (P3) tasks into a single training pipeline, where each mini-batch is sampled from a mixed dataset containing examples of all tasks. 
To further leverage ICL, we incorporate a small number of demonstration pairs into each sequence, directly illustrating the desired relationship between the input and the output. 
These prompts guide the network in correctly interpreting the task, allowing it to focus on relevant features and adapt its predictions.

For precoding tasks P1 and P2, we specify the loss function $\ell(\cdot, \cdot)$ in Eq.~\eqref{eq:icl_objective} to be the Mean Squared Error (MSE) between the predicted and ground-truth low-dimensional parameter vectors that determine the optimal precoder, i.e.,
\begin{equation}
    \ell_{\text{P1}} = \| \hat{\boldsymbol{\lambda}} - \boldsymbol{\lambda}_{\text{gt}} \|_2^2 + \| \hat{\mathbf{p}}- \mathbf{p}_{\text{gt}} \|_2^2,
\end{equation}
\begin{equation}
    \ell_{\text{P2}} = \| \hat{\mathbf{q}} - \mathbf{q}_{\text{gt}} \|_2^2.
\end{equation}
For prediction task P3, the NMSE loss measures the relative error between the predicted channel matrix $\hat{\mathbf{H}}$ and the true channel matrix $\mathbf{H}_{\text{gt}}$, i.e.,
\begin{equation}
    \ell_{\text{P3}} = \frac{ \| \hat{\mathbf{H}} - \mathbf{H}_{\text{gt}} \|_F^2}{\| \mathbf{H}_{\text{gt}} \|_F^2}.
\end{equation}

Finally, the overall training objective $\mathcal{L}_{\text{total}}$ is formulated as a weighted sum of the individual task losses:
\[\mathcal{L}_{\text{total}} = \sum_{t \in \{\text{P1, P2, P3}\}} w_t \mathcal{L}_t\]
where $w_t$ represents the specified weight for task $t$. 

\section{Simulation Results}\label{sec:simulations}
\subsection{Simulation Setup}
\subsubsection{Data Generation}
To generate substantial channel data representing realistic wireless environments,
we utilize the QuaDRiGa channel generator~\cite{Jaeckel2014QuaDRiGa}. 
QuaDRiGa is a well-established tool specifically designed to implement the 3D channel models defined by 3GPP\@.
Our simulation parameters are carefully chosen to align with this standard~\cite{3gpp901} and relevant recent work~\cite{LLM4CP, Multi-taskLLM}.
We specifically adopt the 3GPP Urban Macro (UMa) channel model and focus on non-line-of-sight (NLOS) scenarios.
The channel characteristics are configured with 21 clusters and 20 paths per cluster.
We consider a single-cell downlink MISO-OFDM system. 
The center frequency of the channel is set to 2.4 GHz. 
The bandwidth of the channel is 8.64 MHz, comprising of $M = 48$ subcarriers, i.e., the frequency interval of subcarriers is 180 kHz.
To be consistent with the overviewed literature, only one subcarrier is considered in a time in the input and output of the prediction.
The BS is equipped with a UPA with $N_h=4$ and $N_v=4$ while each user is equipped with a single omnidirectional antenna.
The antenna spacing is half of the wavelength at the center frequency.
The system simultaneously serves $K=4$ moving users. 
User initial positions are randomized, and their motion trajectories are set as linear.
The user velocities are uniformly distributed between 10 and 100 km/h.
The training dataset and test dataset respectively contain 80000 and 10000 samples for each task.
Without loss of generality, we set the power constraint $P_{\text{max}}=1$ in all precoding tasks.
For both precoding tasks, the priority weights $\alpha_k$ and $\rho_k$ are set to 1 for all users, indicating equal importance.
To generate the ground truth labels for the multi-user precoding tasks, we applied high-performance iterative algorithms to each channel sample in the dataset. 
For the sum-rate maximization task (P1), we employed the widely-adopted WMMSE algorithm~\cite{wmmse}, which is known for its effectiveness in achieving near-optimal solutions.
For the max-min SINR task (P2), we implemented the Schubert-Boche algorithm~\cite{Schubert-Boche}, which is proven to converge to the global optimum. 
These algorithms were run for every channel realization until convergence, and the resulting precoding parameters for each task served as the supervised ground truth label for training the ICWLM\@.

\subsubsection{Network and Training Parameters}

\begin{figure}
    \centering
    \includegraphics[width=1\linewidth]{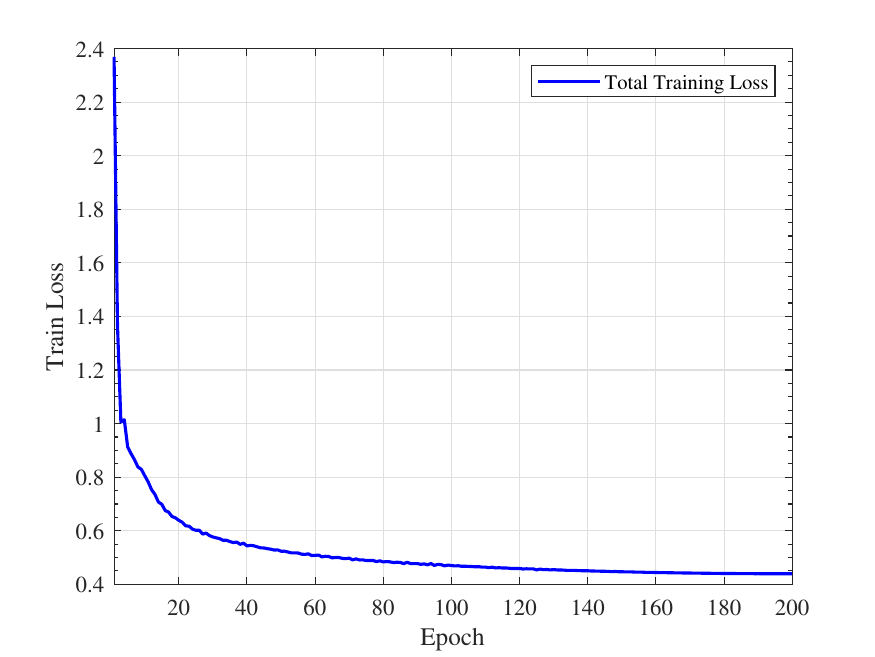}
    \caption{Model Training Loss}\label{fig:training_loss}
\end{figure}

For the proposed ICWLM model, we employ a transformer architecture based on the LLaMA model.
In detail, the model employs 4 layers, 4 attention heads, an embedding dimension of 512, and a FFN dimension of 1024.
The model is trained using the AdamW optimizer.
The optimizer parameters are set to the default PyTorch configuration: $\beta_1=0.9$, $\beta_2=0.999$, and $\epsilon=10^{-8}$, with a weight decay of $10^{-4}$. 
Training is conducted with a batch size of 1024 for 200 epochs.
Both the warm-up and cosine annealing scheduler are employed to train.
In a single training session, the proposed model is jointly trained on both the multi-user precoding and channel prediction datasets. 
Specifically, the precoding dataset encompasses data corresponding to SNR values ranging from 0 dB to 30 dB. 
For fairness, the same total power constraint is enforced on all precoding solutions and signal-to-noise-ratio is defined as $SNR=P_{\text{max}}/ \sigma^2$.
Concurrently, the channel prediction dataset accounts for diverse user velocities, spanning from 10 km/h to 100 km/h. 
For each specific configuration (i.e., each SNR level for precoding tasks and each velocity for the channel prediction task), the dataset comprises 100,000 samples. 
The entire dataset is randomly partitioned into training, validation, and testing sets with a ratio of 8:1:1. 
This substantial volume of diverse data provides a rich supervision signal, enabling the model to learn shared representations and generalize across these distinct physical layer tasks.
Fig.~\ref{fig:training_loss} illustrates the training performance of our model. The loss curve exhibits a sharp initial decline, demonstrating rapid feature learning, before gradually stabilizing. 
This trend confirms that the model successfully converged, validating the effectiveness of our training strategy.

\subsection{Multi-Task Performance Analysis}
\subsubsection{Sum-Rate Maximization Task}
We first evaluate the proposed model on the Sum-Rate Maximization problem (P1). 
To investigate the model's generalization capability across varying channel conditions, we utilize a discrete dataset for training, comprising samples only at SNR levels of 0, 10, 20, and 30 dB. 
During inference, we evaluate the performance on a broader set of SNR levels, crucially including intermediate points (e.g., 5, 15, and 25 dB) that were unseen during the training phase.
We fix the number of in-context demonstration pairs at $l=4$. Specifically, the input sequence is constructed by prepending four task-relevant demonstration pairs to the current query channel matrix, guiding the model to identify the optimization objective.

\begin{figure}
    \centering
    \includegraphics[width=1\linewidth]{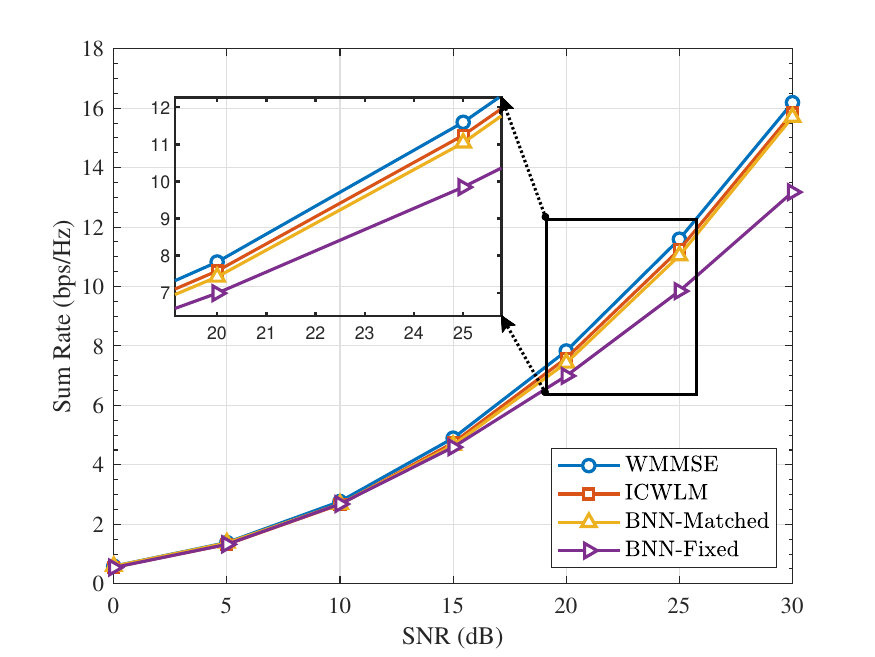}
    \caption{The sum rate performance versus different SNR}\label{fig:o1_fig1}
\end{figure}

For benchmarking, we employ the classical WMMSE algorithm~\cite{wmmse} and a CNN-based Beamforming Neural Network (BNN)~\cite{xia-BF}. 
To rigorously assess adaptability, we analyze two variants of the BNN baseline: \textit{BNN-Matched}, where a separate model is trained and tested specifically for each SNR point, and \textit{BNN-Fixed}, where a single model trained at 10 dB is evaluated across all SNR levels.
The simulation results are plotted in Fig.~\ref{fig:o1_fig1}. 
It can be observed that the proposed model achieves a sum-rate performance comparable to the iterative WMMSE algorithm across the entire evaluated SNR range. 
A key finding is the model's exceptional generalization ability, as evidenced by its sustained high accuracy even on unseen SNR points (5, 15, and 25 dB), which demonstrates effective interpolation capabilities.
In stark contrast, the limitation of conventional small-scale deep learning models is evident in the BNN baselines. 
While the \textit{BNN-Matched} achieves competitive results, it necessitates retraining a dedicated model for each parameter configuration. 
The \textit{BNN-Fixed} model, however, suffers from significant performance degradation when the testing SNR deviates from its training condition (10 dB). 
Unlike these specialized small models that struggle with parameter shifts, our proposed approach effectively handles diverse and unseen SNR conditions using a single pre-trained backbone, highlighting the flexibility and efficiency of the ICL paradigm.

\subsubsection{SINR Balancing Task}
\begin{figure}
    \centering
    \includegraphics[width=1\linewidth]{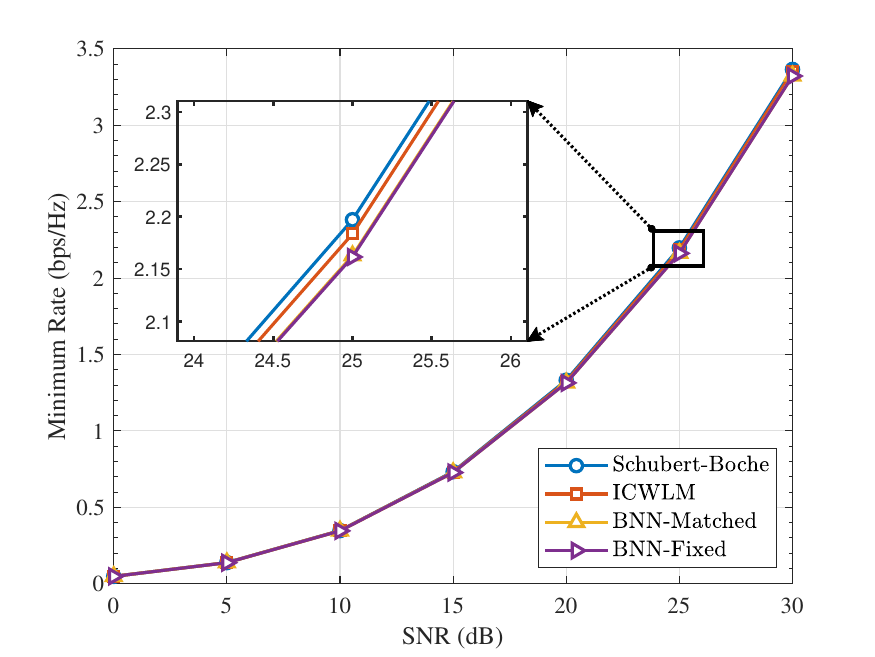}
    \caption{The minimum rate of users versus different SNR}\label{fig:o2_fig1}
\end{figure}

We next turn our attention to the SINR Balancing problem (P2), which aims to maximize the minimum SINR among all users subject to a total power constraint.
To ensure a consistent evaluation framework, we adopt the same experimental configuration as used in the sum-rate maximization task. 
We benchmark our proposed approach against two baselines: the classical Schubert-Boche algorithm~\cite{Schubert-Boche}, which provides the theoretical optimal solution for the max-min SINR problem, 
and the CNN-based BNN model~\cite{xia-BF} (including both Matched and Fixed variants). 
The performance results are presented in Fig.~\ref{fig:o2_fig1}.

Similar to the previous task, our model demonstrates remarkable adaptability. 
When provided with the optimal solutions as in-context examples, it effectively learns the mapping function and achieves performance close to the theoretical optimal baseline.
Unlike the sum-rate task, we observe that the BNN-Fixed and BNN-Matched baselines exhibit comparable performance, indicating that the interference balancing strategy is less sensitive to SNR variations. 
Despite this strong baseline performance, our proposed model consistently maintains high accuracy across all SNR levels, further validating its robust generalization capability across different optimization objectives.

\subsubsection{Channel Prediction Task}
Finally, we evaluate the performance of the proposed model on the Channel Prediction task (P3), which aims to forecast the CSI of the target time slot based on historical channel sequences.
To rigorously assess the model's generalization capability across different mobility profiles, we adopt a discrete training strategy. 
The model is trained using datasets generated at specific user velocities of $\{10, 40, 70, 100\}$ km/h. 
During the inference phase, however, the evaluation covers a comprehensive velocity range from 10 km/h to 100 km/h, crucially including intermediate velocities (e.g., 20, 30, 50 km/h) that were unseen during the training phase.
For the in-context inference setup, we fix the number of demonstration pairs at $l=4$. 
Given the overlapping data format of this task, this configuration effectively provides the model with a context window comprising the CSI from the past 5 historical time slots to guide the prediction of the next instance.

\begin{figure}
    \centering
    \includegraphics[width=1\linewidth]{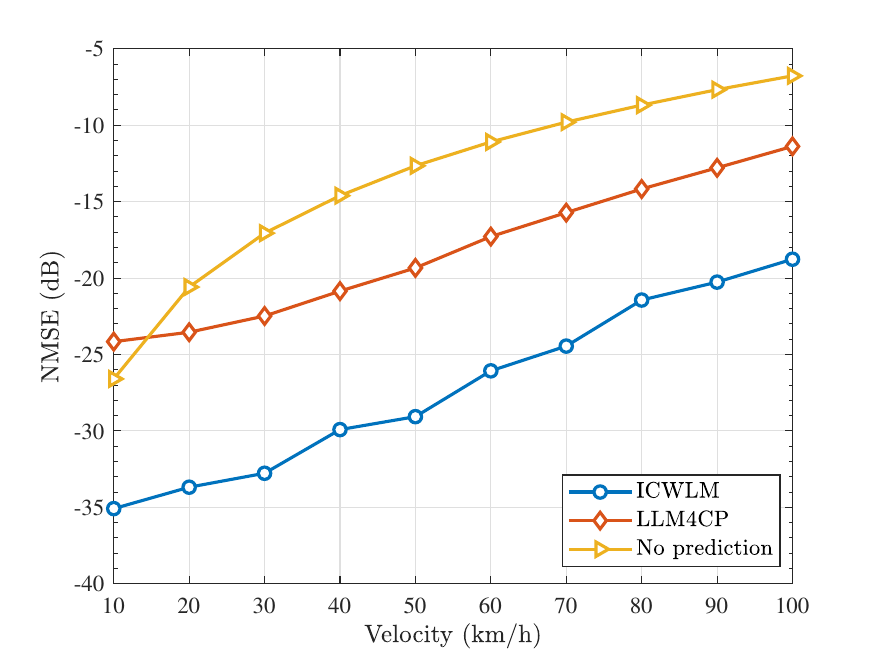}
    \caption{The NMSE performance versus different user velocities}\label{fig:o3_fig1}
\end{figure}

For benchmarking, we compare our approach against two distinct baselines. 
First, we employ the LLM4CP model~\cite{LLM4CP}, which represents a state-of-the-art llm-based channel predictor. 
Second, to quantify the impact of channel aging, we include the Outdated CSI scheme (also referred to as ``No Prediction''). 
This baseline directly utilizes the CSI from the most recent time slot as the prediction for the future, serving as a performance lower bound.
The NMSE performance against varying user velocities is presented in Fig.~\ref{fig:o3_fig1}. 
Intuitively, as user mobility increases, the temporal correlation of CSI significantly decreases due to higher Doppler shifts, making accurate prediction inherently more challenging. 
This is evident in the sharp performance degradation of the Outdated CSI baseline.
However, our proposed ICWLM model consistently achieves significantly lower NMSE compared to both baselines across the entire velocity spectrum.
A pivotal finding is the model's robustness on the unseen velocity points. 
Despite only being trained on a sparse set of mobility patterns, the model demonstrates remarkable interpolation capabilities, maintaining high prediction accuracy even at unobserved velocities. 
This insensitivity to velocity variations validates the superior generalization capabilities of our model in dynamic wireless environments, ensuring reliable channel knowledge for subsequent physical layer operations.

\subsection{In-context Learning Ability}

To evaluate the in-context learning capability of our proposed ICWLM model, we investigate the impact of context size by varying the number of demonstration pairs during inference. 
We randomly select a subset of demonstration pairs from the dataset as prompts.
In the following results, the notation ``ICL-$l$'' denotes the inference setting where $l$ demonstration pairs are provided as the context.

\begin{figure}
    \centering
    \includegraphics[width=1\linewidth]{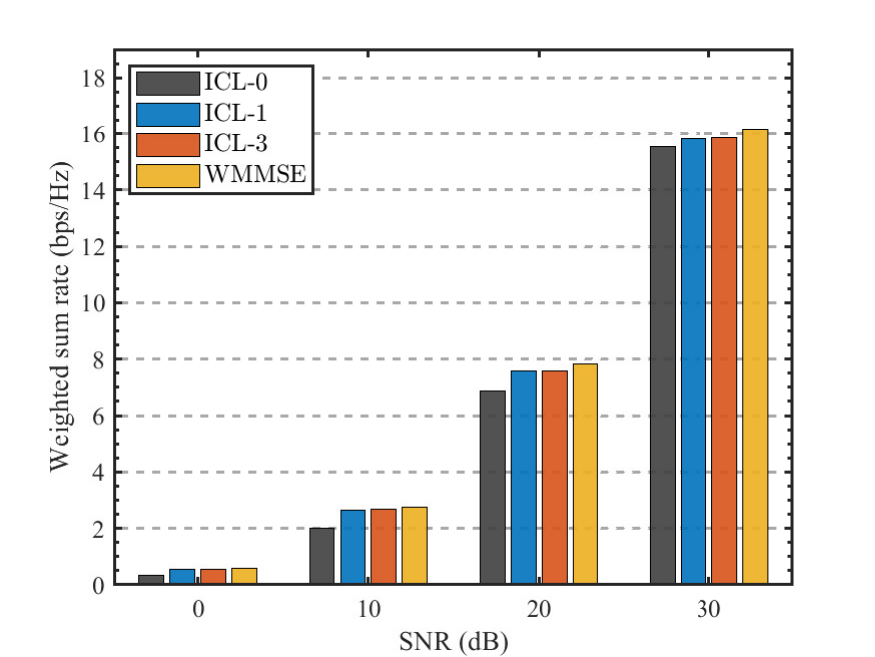}
    \caption{The ICL performance of the proposed model on P1}\label{fig:o1_fig2}
\end{figure}

\begin{figure}
    \centering
    \includegraphics[width=1\linewidth]{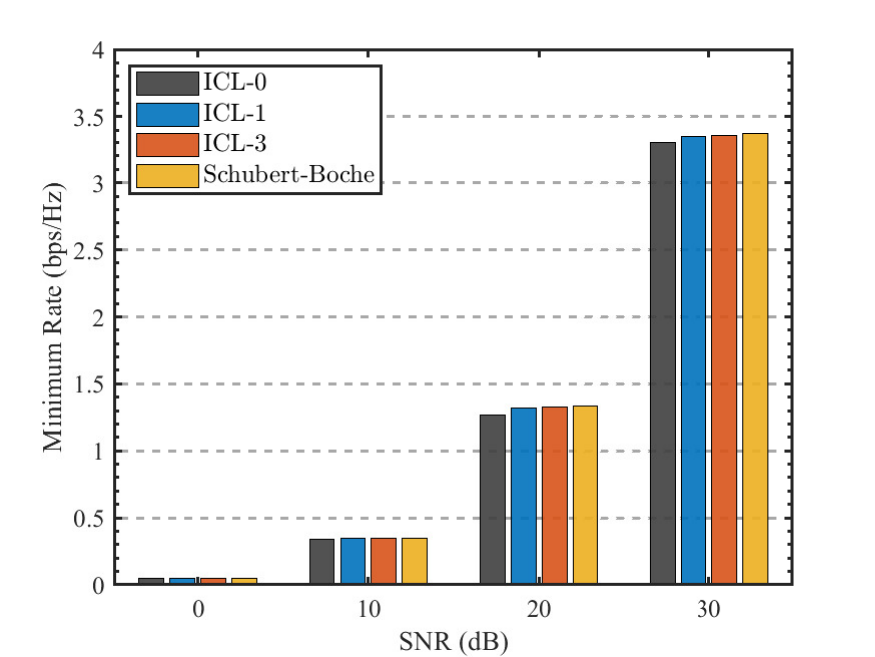}
    \caption{The ICL performance of the proposed model on P2}\label{fig:o2_fig2}
\end{figure}

\begin{figure}
    \centering
    \includegraphics[width=1\linewidth]{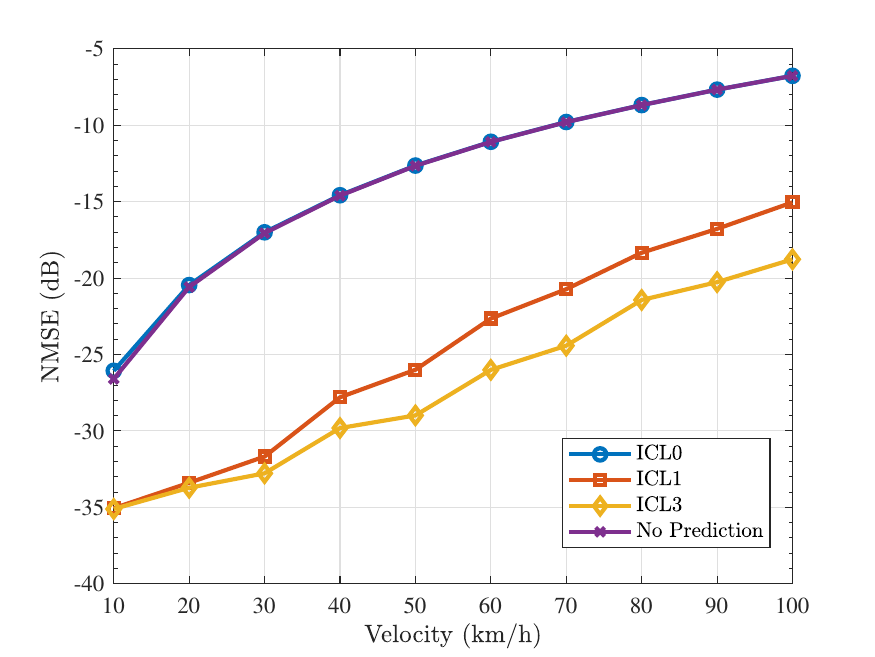}
    \caption{The ICL performance of the proposed model on P3}\label{fig:o3_fig2}
\end{figure}

For the Sum-Rate Maximization task (P1), as shown in Fig.~\ref{fig:o1_fig2}, the zero-shot setting (ICL-0) struggles to achieve optimal performance. 
However, providing just a single demonstration pair (ICL-1) yields a substantial performance boost, bringing the weighted sum rate remarkably close to the iterative WMMSE baseline. 
Increasing the context size to three examples (ICL-3) results in marginal further improvements, indicating that the model can rapidly grasp the optimization objective with minimal supervision.
For the SINR Balancing task (P2), the results in Fig.~\ref{fig:o2_fig2} exhibit a similar trend. 
While ICL-0 performs adequately at low SNR, it falls short of the optimal solution as SNR increases. 
The introduction of one-shot learning (ICL-1) effectively bridges this gap, enabling the model to match the performance of the theoretical optimal Schubert-Boche algorithm. 
This demonstrates the model's ability to identify the max-min fairness constraint efficiently from limited examples.
For the Channel Prediction task (P3), Fig.~\ref{fig:o3_fig2} illustrates the NMSE performance against user velocity. 
The ``No Prediction'' baseline serves as a lower bound, showing severe degradation at high speeds. 
Consistent with the precoding tasks, ICL-0 provides a baseline prediction capability but is outperformed by in-context learning settings. 
Notably, ICL-1 significantly reduces the prediction error across all velocities, and ICL-3 offers further refinement, demonstrating that additional temporal context helps the model better capture the channel evolution dynamics in high-mobility scenarios.

\section{Conclusions}\label{sec:conclusion}
In this paper, we proposed a multi-task wireless large model called ICWLM that can handle multiple physical layer tasks through in-context learning.
Specifically, we first formulated two fundamental problems in wireless communications: multi-user precoding and channel prediction. 
Then, we designed a transformer-based architecture that can process different types of wireless data in a unified framework. 
The proposed model leverages the in-context learning capability to distinguish and solve different tasks with only a few demonstration pairs as prompts.
Extensive simulation results demonstrated that our ICWLM can achieve competitive performance compared with task-specific methods 
while maintaining good generalization ability for unseen scenarios. 
This work provides a new perspective on developing unified AI models for wireless communications, 
which can potentially reduce the deployment complexity of intelligent networks.
Future work will focus on extending ICWLM to a broader range of wireless tasks, enhancing its computational efficiency for edge deployment through model compression, and validating its real-world performance and robustness on hardware testbeds.

\bibliographystyle{IEEEtran}
\bibliography{icwlm, IEEEabrv}

\end{document}